\documentclass[twocolumn,showpacs,preprintnumbers,amsmath,amssymb,prb]{revtex4}

\usepackage{graphicx}
\usepackage{dcolumn}
\usepackage{bm}

\begin{document}


\title{Spin and density excitations in the triangular-lattice 
$t$-$J$ model with multiple-spin exchange interactions: 
$^3$He on graphite}

\author{K. Seki,$^1$ T. Shirakawa,$^{1,2}$ and Y. Ohta$^1$}
\affiliation{$^1$Department of Physics, Chiba University, 
Chiba 263-8522, Japan}
\affiliation{$^2$Institut f\"ur Theoretische Physik, 
Leibniz Universit\"at Hannover, D-30167 Hannover, Germany}

\date{\today}

\begin{abstract}
Using an exact diagonalization technique on small clusters, we 
study spin and density excitations of the triangular-lattice 
$t$-$J$ model with multiple-spin exchange interactions, whereby 
we consider anomalous properties observed in the doped Mott 
region of the two-dimensional liquid $^3$He adsorbed on a 
graphite surface.  
We find that the double-peak structure consistent with experiment 
appears in the calculated temperature dependence of the specific 
heat; the low-temperature sharp peak comes from the spin 
excitations reflecting the frustrated nature of the spin 
degrees of freedom and high-temperature broad peak comes from 
the density excitations extending over the entire band width.  
The clear separation in their energy scales is evident in the 
calculated spin and density excitation spectra.  
The calculated single-particle excitation spectra suggest the 
presence of fermionic quasiparticles dressed by the spin 
excitations, with an enhanced effective mass consistent with 
experiment.  
\end{abstract}


\maketitle

\section{Introduction}

$^3$He atoms adsorbed on a graphite surface is known to be an 
ideal two-dimensional correlated spin-1/2 fermion system.  
A solidified commensurate phase of $^3$He atoms is stabilized 
at a $4/7$ density of the underlying layer of $^4$He atoms due to 
the substrate potential corrugation and thus a triangular lattice 
of $^3$He atoms is formed, which is a realization of a gapless 
quantum spin liquid (QSL).\cite{Fukuyama}  
Theoretically, this $4/7$ phase of spin-1/2 $^3$He atoms has been 
studied by using the triangular-lattice Heisenberg model with 
the multiple-spin exchange 
interactions.\cite{Roger1,Roger2,Misguich,Momoi1,Momoi2,Bauerle}  
Importance of the density fluctuations has recently been pointed 
out as well.\cite{Watanabe}  

A finite amount of vacancies of $^3$He atoms can be introduced 
into this $4/7$ phase in a stable manner, where the vacancies 
can hop from site to site of the triangular lattice via 
quantum-mechanical tunneling motions even at absolute zero 
temperature.  
The presence of such vacancies, called the zero-point vacancies 
(ZPVs), was predicted a few decays ago.\cite{Andreev1,Matsuda}  
Quite recently, the experimental evidence for the ZPVs has 
been reported in the monolayer of $^3$He adsorbed on a surface 
of graphite preplated by a solid monolayer of 
$^4$He:\cite{Matsumoto1,Matsumoto2} i.e., heat capacity 
measurements of the system show an anomalous coexistence of 
a magnetic round-peak near 1 mK and a broad peak at several 
tens mK that are associated with the ZPVs doped into the 
commensurate Mott-localized solid.  The ZPVs are maintained 
up to the doping of almost 20\% of the lattice sites, which we 
call the doped Mott region of monolayer $^3$He.  

Theoretically, Fuseya and Ogata \cite{Fuseya} have proposed 
the triangular-lattice $t$-$J$ model with four-spin 
ring-exchange interactions as an effective model for the 
doped Mott region of the system and obtained its ground-state 
phase diagram.  The low-energy excitations of the model were 
also discussed.  They have thereby argued that there is a 
new-type anomalous quantum-liquid phase, characteristic 
of the ``spin-charge separation'', which may be relevant with 
the anomalous features observed in the doped Mott region 
of the monolayer $^3$He adsorbed on a graphite surface.  

Motivated by such developments in the field, we study in 
this paper the triangular-lattice $t$-$J$ model with the 
multiple-spin exchange interactions further.  In particular, 
we directly calculate the spin and density excitation 
spectra and single-particle spectra as well as the 
temperature dependence of the specific heat and uniform 
magnetic susceptibility by using an exact-diagonalization 
technique on small clusters.  
We thereby consider the anomalous properties observed in the 
doped Mott region of the two-dimensional liquid $^3$He 
adsorbed on a graphite surface.  

We will thus demonstrate that the double-peak structure actually 
appears in the temperature dependence of the specific heat, 
which is quantitatively consistent with experiment; the 
low-temperature sharp peak comes from the spin excitations 
and high-temperature broad peak comes from the density 
excitations.  
The spectral weight for the calculated spin excitation spectra 
is concentrated on a very low-energy region that scales with 
the exchange interactions, while that of the density 
excitations extends over an entire band width that scales 
with the hopping parameter of the vacancy.  The clear separation 
between spin and density excitations in their energy scales is 
thus found.  The accumulation of the low-energy spectral weight 
of the spin excitations comes from the frustrated nature of 
the spin degrees of freedom of the system; i.e., the 
ferromagnetic two-spin interactions compete with the 
antiferromagnetic four-spin interactions on the geometrically 
frustrated triangular lattice.  
The single-particle excitation spectra suggest that the 
vacancies behave like fermionic quasiparticles dressed by the 
spin excitations, with the enhanced effective mass consistent 
with experiment.  
Preliminary results of our work have been presented 
in Ref.~\cite{Seki}.  

This paper is organized as follows. 
In Sec.~II, we present our model and method of calculation.  
In Sec.~III, we present our results of calculations for 
the specific heat, magnetic susceptibility, spin and density 
excitation spectra, and single-particle excitation spectra.  
We compare our results with experiment in Sec.~IV.  
We summarize our work in Sec.~V.  

\begin{figure}[th]
\begin{center}
\resizebox{6.0cm}{!}{\includegraphics{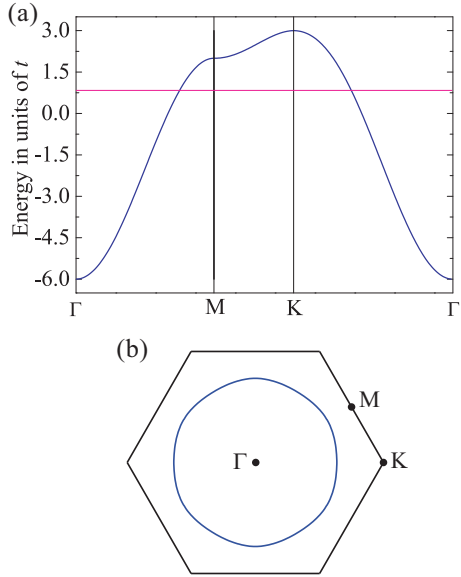}}\\
\caption{(Color online) (a) Noninteracting tight-binding band 
structure of the triangular lattice with the nearest-neighbor 
hopping parameter $t$.  Horizontal line represents the Fermi 
energy at half filling.  (b) Brillouin zone and Fermi surface 
at half filling.}
\label{band}
\end{center}
\end{figure}

\section{Model and method}

The triangular-lattice $t$-$J$ model with the multiple-spin exchange 
interactions is defined by the Hamiltonian
\begin{eqnarray}
{\cal H}&=&-t\sum_{\langle ij\rangle,\sigma}
\big({\tilde c}_{i\sigma}^{\dagger}{\tilde c}_{j\sigma}+{\rm H.c.}\big)
+J\sum_{\langle ij\rangle}
\Big({\bf S}_i\cdot{\bf S}_j-\frac{n_in_j}{4}\Big)\cr
&&+K\sum_{\langle ijkl\rangle}(P_4+P_4^{-1})
+R\sum_{\langle ijklmn\rangle}(P_6+P_6^{-1})
\end{eqnarray}
where ${\tilde c}_{i\sigma}=c_{i\sigma}(1-n_{i,-\sigma})$ is 
the projected annihilation operator of a fermion ($^3$He atom) 
at site $i$ and spin $\sigma$ $(=\uparrow,\downarrow))$ 
allowing no doubly occupied sites, ${\bf S}_i$ is the spin-1/2 
operator, and $n_i$ $(=n_{i\uparrow}+n_{i\downarrow})$ is the 
number operator.  
The summation in the $t$-$J$ part of the model is taken over 
all the nearest-neighbor pairs $\langle ij\rangle$ on the 
triangular lattice.  
$P_4$ and $P_6$ are the four-spin and six-spin exchange operators 
defined as $P_4=P_{il}P_{ik}P_{ij}$ and 
$P_6=P_{in}P_{im}P_{il}P_{ik}P_{ij}$, respectively, where 
$P_{ij}=(1+{\bm\sigma}_i\cdot{\bm\sigma}_j)/2$ with 
the Pauli spin matrix ${\bm\sigma}_i$.  
The summation is taken over all the possible combinations 
of four nearest-neighbor sites $\langle ijkl\rangle$ for 
$P_4$ and over all the equilateral hexagons 
$\langle ijklmn\rangle$ for $P_6$.  

In this paper, we study the dynamical properties of the 
model under the introduction of vacancies in the 4/7 commensurate 
solid phase of $^3$He, i.e., removal of particles ($^3$He atoms) 
or addition of ZPVs.  We thus define the filling $n$ of particles 
as $n=N/L$ where $N$ is the total number of particles and $L$ is 
the total number of lattice sites in the system; in particular, 
$n=1$ is referred to as ``half filling'', which corresponds to 
the 4/7 solid phase.  
The noninteracting band structure and Fermi surface at half 
filling of the tight-binding model with the nearest-neighbor 
hopping parameter $t$ are shown in Fig.~1; 
we find no nesting features in the Fermi surface and no 
singularities in the density of states for $n<1$.  

The nearest-neighbor hopping parameter $t$ and two-spin and 
four-spin exchange interaction parameters $J$ and $K$ have been 
estimated as follows:\cite{Fuseya} $t\simeq 50-100$ mK, 
$-J\simeq 1-10$ mK, and $K/|J|\sim 0.2$. 
We note that, in this parameter region, the two-spin exchange 
term favors the ferromagnetic spin polarization ($J<0$) 
but the four-spin exchange term gives the antiferromagnetic 
spin correlations between neighboring spins, and thus we have 
the situation where the strong frustration in the spin 
degrees of freedom of the system appears.  Geometrical 
frustration also appears on the triangular lattice when 
the interaction between spins is antiferromagnetic.  
We have examined the effects of the six-spin exchange 
interaction term on the ground state and excitation spectra 
and found that the effects are very small, in particular 
when doped with vacancies.  This is because the $P_6$ exchange 
interaction is easily cut by the presence of vacancies.  
We will therefore present the results at $R=0$ in this paper, 
the model of which we refer to as the $t$-$J$-$K$ model.  

Throughout the paper, we use $t=1$ as the unit of energy 
unless otherwise stated and we set $\hbar=k_{\rm B}=1$.  

\begin{figure}[th]
\begin{center}
\resizebox{7.5cm}{!}{\includegraphics{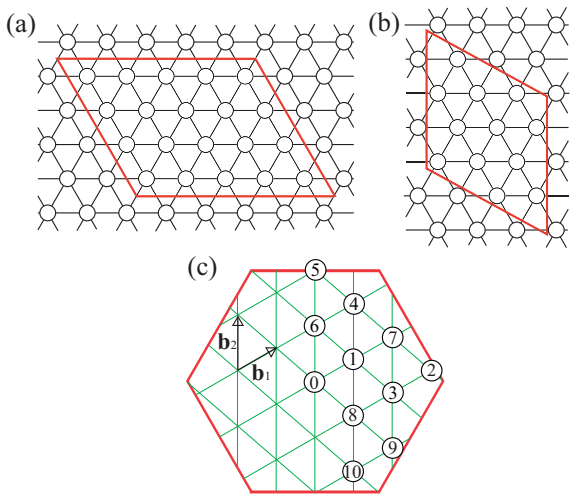}}\\
\caption{(Color online) (a) 20-site and (b) 12-site clusters 
used for calculations.  (c) Brillouin zone and available 
momenta of the 20-site cluster in the periodic boundary 
condition.}
\label{cluster}
\end{center}
\end{figure}

We use the Lanczos exact-diagonalization technique on 
small clusters to calculate the ground state and excitation 
spectra of the model.  
In particular, we calculate the dynamical spin and density 
correlation functions defined, respectively, as 
\begin{eqnarray}
S({\bf q},\omega)=-\frac{1}{\pi}\Im 
\langle\Psi_0|
{S_{\bf q}^z}^\dagger
\frac{1}{\omega+i\eta-({\cal H}-E_0)}
S_{\bf q}^z
|\Psi_0\rangle
\end{eqnarray}
and 
\begin{eqnarray}
N({\bf q},\omega)=-\frac{1}{\pi}\Im 
\langle\Psi_0|
n_{\bf q}^\dagger
\frac{1}{\omega+i\eta-({\cal H}-E_0)}
n_{\bf q}
|\Psi_0\rangle ,
\end{eqnarray}
where $\Psi_0$ and $E_0$ are the ground-state wave 
function and energy, respectively.  $S_{\bf q}^z$ 
and $n_{\bf q}$ are the Fourier transforms of the 
spin and particle-number operators defined, 
respectively, as 
\begin{eqnarray}
S_{\bf q}^z&=&\frac{1}{\sqrt{L}}\sum_i
e^{i{\bf q}\cdot{\bf r}_i}S_i^z\\
n_{\bf q}&=&\frac{1}{\sqrt{L}}\sum_i
e^{i{\bf q}\cdot{\bf r}_i}n_i ,
\end{eqnarray}
where ${\bf r}_i$ is the position of the lattice site $i$.  

We also calculate the single-particle excitation 
spectrum defined as 
\begin{eqnarray}
A({\bf q},\omega)=A^-({\bf q},-\omega)+A^+({\bf q},\omega)
\end{eqnarray}
with the particle removal spectrum 
\begin{eqnarray}
A^-({\bf q},\omega)=-\frac{1}{\pi}\Im 
\langle\Psi_0|
{\tilde c}_{{\bf q}\sigma}^\dagger
\frac{1}{\omega+i\eta-({\cal H}-E_0)}
{\tilde c}_{{\bf q}\sigma}
|\Psi_0\rangle
\end{eqnarray}
and particle addition spectrum 
\begin{eqnarray}
A^+({\bf q},\omega)=-\frac{1}{\pi}\Im 
\langle\Psi_0|
{\tilde c}_{{\bf q}\sigma}
\frac{1}{\omega+i\eta-({\cal H}-E_0)}
{\tilde c}_{{\bf q}\sigma}^\dagger
|\Psi_0\rangle ,
\end{eqnarray}
where $\eta\rightarrow +0$, which is replaced by a 
small positive number in the actual calculations 
to give an artificial broadening of the spectra.  
We use a cluster of 20 sites with periodic boundary 
condition for these calculations (see Fig.~2), where 
the independent available momenta in the Brillouin zone, 
${\bf q}_0$, $\cdots$, ${\bf q}_{10}$, are also shown.  
They are at ${\bf q}_i=m{\bf b}_1+n{\bf b}_2$ with 
${\bf b}_1=2\pi/5(1,1/\sqrt{3})$ and 
${\bf b}_2=(0,\pi/\sqrt{3})$, where $(m,n)$ are 
$(0,0)$ for ${\bf q}_0$, 
$(1,0)$ for ${\bf q}_1$, 
$(3,-1)$ for ${\bf q}_2$, 
$(2,-1)$ for ${\bf q}_3$, 
$(1,1)$ for ${\bf q}_4$, 
$(0,2)$ for ${\bf q}_5$, 
$(0,1)$ for ${\bf q}_6$, 
$(2,0)$ for ${\bf q}_7$, 
$(1,-1)$ for ${\bf q}_8$, 
$(2,-2)$ for ${\bf q}_9$, and 
$(1,-2)$ for ${\bf q}_{10}$.  
In the following, we will in particular examine the 
cluster with two vacancies, i.e., $n=0.9$.  

To calculate the temperature $T$ dependence of the specific 
heat $C(T)$, magnetization $M(T)$ under uniform magnetic 
field $h$, and uniform magnetic susceptibility 
$\chi(T)=\lim_{h\rightarrow 0}(\partial M\big/\partial h)_T$, 
the Hamiltonian for a smaller-size cluster of 12 sites 
(see Fig.~2) is fully diagonalized to calculate the partition 
function.\cite{Shannon,Elstner}  We add the Zeeman term 
$-h\sum_iS_i^z$ to the Hamiltonian Eq.~(1) when we 
calculate the magnetic response of the system.  
In the following, we will in particular examine this cluster 
with one vacancy ($n=0.92$) and two vacancies ($n=0.83$).

\section{Results of calculation}

\subsection{Ground-state phase diagram}

The ground-state phase diagram in the parameter space of the 
present model at $R=0$ has been obtained by Fuseya and 
Ogata,\cite{Fuseya} which we have also reproduced successfully.  
Here, we briefly review their results.  The phases obtained 
are as follows: 
the region of phase separation (phase-I), 
the region of Fermi liquid with strong spin fluctuations (phase-II), 
the region of new-type anomalous quantum liquid (phase-III), 
and the region of ferromagnetism (phase-IV), where we follow 
their notations of the phases.  
They have put special emphasis on the phase-III, which has 
been argued to be the region of ``spin-charge separation'' 
and may be relevant with the anomalous features of the doped 
Mott region of the $^3$He monolayer.  

In the following, we in particular examine the region of this 
new-type anomalous quantum liquid (phase-III) using the parameter 
values $J=-0.3$ and $K=0.06$, which we compare with the results 
of other regions when necessary, i.e., the region of 
ferromagnetism (phase-IV) using $J=-0.3$ and $K=0$ and 
the region of Fermi liquid (phase-II) using $J=-0.3$ and 
$K=0.15-0.2$.  

\begin{figure}[th]
\begin{center}
\resizebox{8.5cm}{!}{\includegraphics{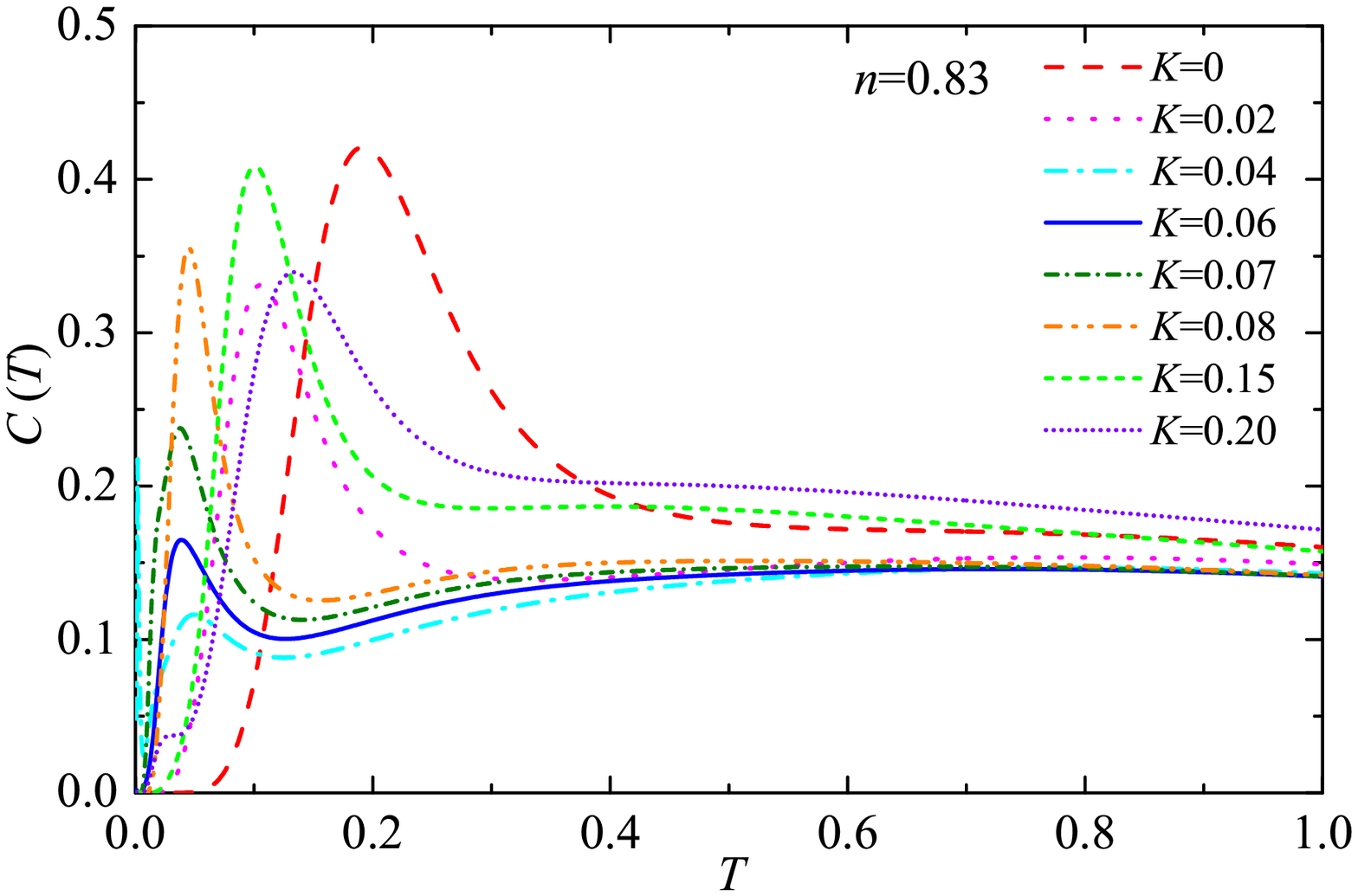}}\\
\medskip
\resizebox{8.5cm}{!}{\includegraphics{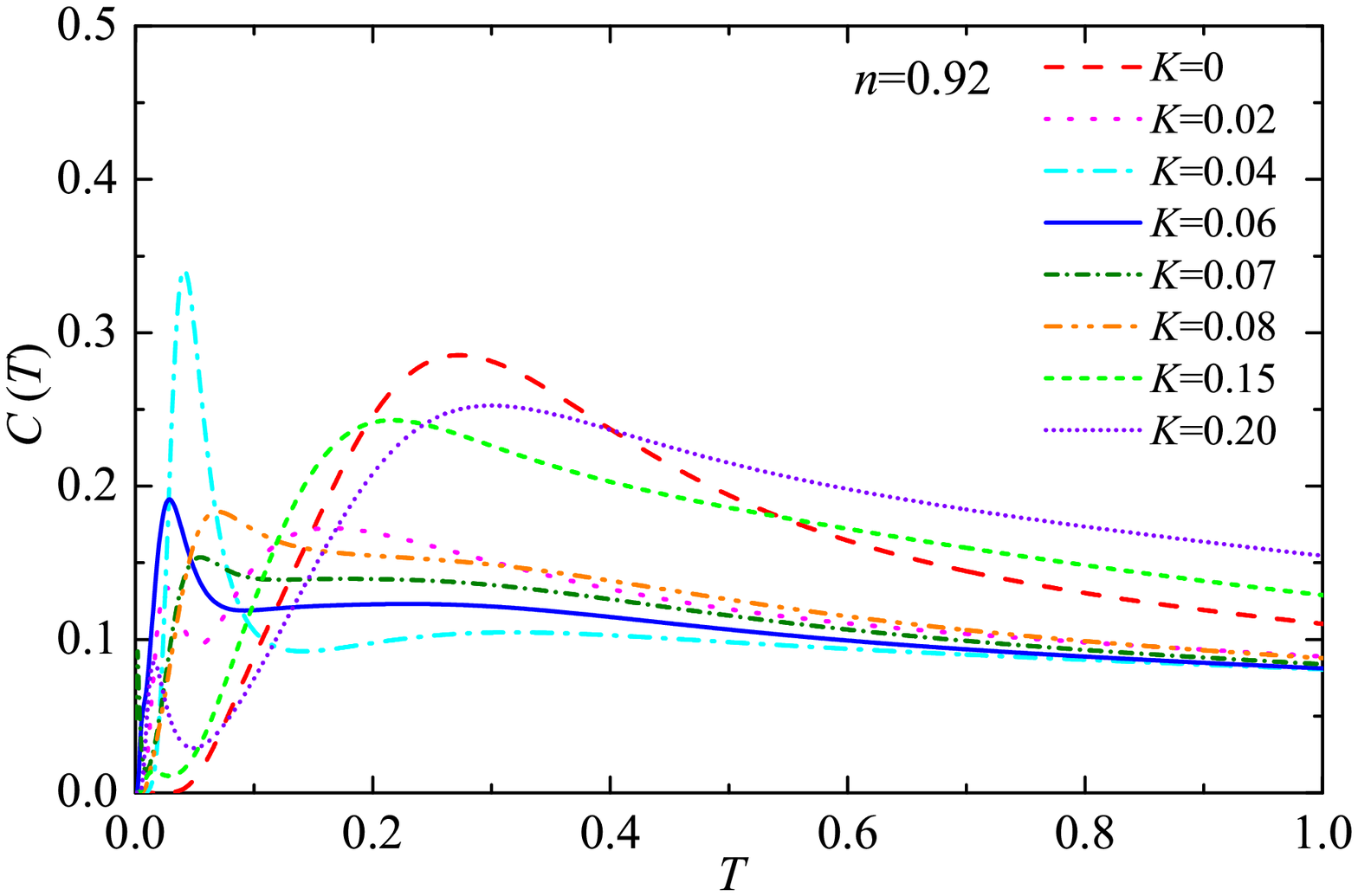}}\\
\caption{(Color online) Calculated temperature dependence of 
the specific heat (per site) of the $t$-$J$-$K$ model at 
$J=-0.3$.  We use the cluster of 12 sites with two vacancies 
($n=0.83$) in the upper panel and with one vacancy ($n=0.92$) 
in the lower panel.}
\label{specificheat}
\end{center}
\end{figure}

\subsection{Specific heat and entropy}

The calculated results for the temperature dependence of 
the specific heat $C(T)$ at $n=0.83$ and $0.92$ are shown 
in Fig.~3.  We find that there appears a double-peak structure 
in $C(T)$ at $K=0.06$ (corresponding to the new-type anomalous 
quantum-liquid phase); i.e., a sharp peak at low temperatures 
and a very broad peak extending over high temperatures.  
The double-peak structure is not clearly seen at $K=0$ and 
$0.15-0.2$.  
We should note here that the specific heat coefficient 
$\gamma$, where $C(T)=\gamma T$ at low temperatures, 
cannot be deduced from the present calculations since 
$C(T)$ decays exponentially at low temperatures due to the 
discreteness of the energies of finite-size systems.  

We will show in Sec.~III D that the low-temperature sharp 
peak comes from the excitation of the spin degrees of freedom 
of the system and the broad high-temperature peak comes from 
the excitations of the density degrees of freedom of the 
system.  In other words, the width of the sharp low-energy 
peak scales with the exchange interactions between spins 
(a combination of $J$ and $K$) and the width of the broad 
high-energy peak scales with the hopping parameter $t$ of 
the vacancy.  

It is interesting to note that the double-peak structure in 
the specific heat $C(T)$ due to the separation in their 
energy scales between spin and density degrees of freedom has 
previously been discussed in the context of the low-energy 
excitations in the Hubbard ladder systems with charge ordering 
instability although the latter is for the insulating systems 
with a charge gap.\cite{Ohta}  
In Sec.~IV, we will compare the obtained double-peak 
structure with experiment\cite{Matsumoto1,Matsumoto2}.  
We also calculate the temperature dependence of 
the entropy $S(T)$ (not shown here), which will be compared 
with experiment\cite{Matsumoto2} also in Sec.~IV.  

\begin{figure}[th]
\begin{center}
\resizebox{7.0cm}{!}{\includegraphics{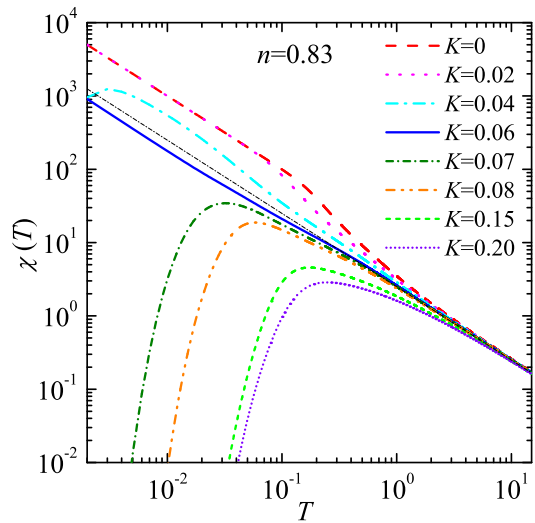}}\\
\caption{(Color online) Calculated temperature dependence of 
the uniform magnetic susceptibility of the $t$-$J$-$K$ 
model at $J=-0.3$.  We use the cluster of 12 sites with 
two vacancies: $n=0.83$.  Straight line corresponds to 
the Curie law, $\chi(T)=C/T$, where $C$ is the Curie 
constant.}
\label{susceptibility}
\end{center}
\end{figure}

\subsection{Uniform magnetic susceptibility}

The calculated results for the temperature dependence of 
the uniform magnetic susceptibility $\chi(T)$ are shown in 
Fig.~4.  
We find that the temperature variation is strongly 
dependent on the value of $K$: 
(i) When $0\le K\lesssim 0.04$, $\chi(T)$ is strongly 
enhanced in comparison with the Curie susceptibility 
$\chi(T)=C/T$, resulting in the ferromagnetic spin 
polarization at low temperatures.  
(ii) When $K=0.06$ at which the frustration in the spin 
degrees of freedom is the largest, $\chi(T)$ is slightly 
suppressed in comparison with the Curie law.  Here, the 
ground state is highly degenerate due to the frustration 
of the spin degrees of freedom; in our cluster, the 
degeneracy is 15 fold.  
(iii) When $K\gtrsim 0.07$, $\chi(T)$ is rapidly 
suppressed with decreasing temperatures.  Here, the 
ground state of the system is spin singlet without 
degeneracy.  

It should be noted that these results come basically from 
the finite-size effects of small clusters.  However, we 
may infer the intrinsic nature of the infinite-size system 
and its $K$ dependence from the low-energy behavior under 
the magnetic field.  
The results for $\chi(T)$ thus obtained are compared 
with experiment in Sec.~IV.  

\begin{figure}[th]
\begin{center}
\resizebox{7.0cm}{!}{\includegraphics{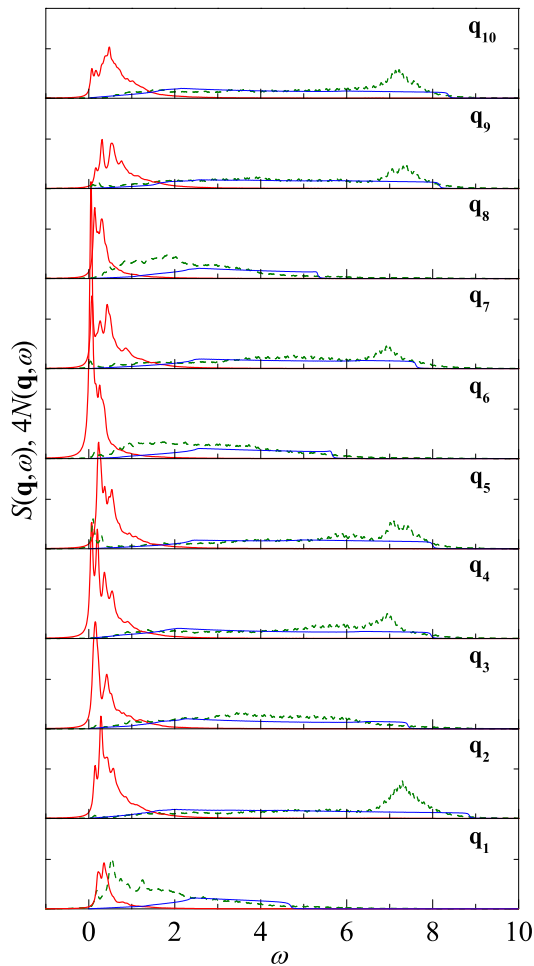}}\\
\caption{(Color online) Calculated spin (solid line) and 
density (dashed line) excitation spectra of the 
$t$-$J$-$K$ model at $J=-0.3$ and $K=0.06$.  
$\eta=0.04$ is assumed.  We use the 20-site cluster 
with two vacancies ($n=0.9$).  The spectra for the 
noninteracting infinite system $N_0({\bf q},\omega)$ 
are also shown for comparison (thin solid line).}
\label{spindensityexcitationspectra}
\end{center}
\end{figure}

\subsection{Spin and density excitation spectra}

The calculated results for the spin and density excitation 
spectra at $K=0.06$ are shown in Fig.~5.  
The excitation spectra for the corresponding noninteracting 
infinite-size system, $N_0({\bf q},\omega)=2S_0({\bf q},\omega)$, 
are also shown for comparison.  We find the following.  
The spectral weight for the spin excitations is concentrated 
on a very low-energy region of around $\omega\lesssim 1$.  
This reflects the presence of a large number of nearly degenerate 
low-energy states coming from the frustrated nature of the spin 
degrees of freedom.  
The low-energy spectral weight is extended over the entire 
Brillouin zone, rather than special momenta, reflecting the 
spatially localized nature of the spin fluctuations.  
The spectral weight for the density excitations, on the other 
hand, extends over a wide energy range of about $0<\omega\lesssim 9$ 
(entire band width), which is more or less resembles the 
spectrum of the noninteracting system.  

From these results, we may say that the spin and density 
excitations are clearly separated in their energy scales: 
i.e., the spin excitations concentrate on the low energy regions, 
the width of which scales with the exchange interactions (a 
combination of $J$ and $K$), and the charge excitations extend 
over the entire band width, which scales with the hopping parameter 
$t$ of the vacancy.  With increasing $K$, we find the upward shift 
of the low-energy spectral weight of the spin excitations; e.g., 
at $K=0.15-0.2$, we find the peaks at a higher-energy region 
of around $0<\omega\lesssim 5$, where the momentum dependence 
of the positions of the peaks becomes significant as well.  
Thus, the separation between the energy scales of the spin 
and density excitations becomes weaker.  

We should note that the energy range where the spectral weight 
of the spin excitations accumulates, i.e., $0<\omega\lesssim 0.1$, 
corresponds well to the temperature range where the low-temperature 
sharp peak in the calculated specific heat $C(T)$ appears.  
We should also note that the broad spectra extending over the 
entire band width correspond well to the very broad high-temperature 
peak in the calculated result for $C(T)$.  
We may therefore conclude that the separation between the spin 
and density excitations in their energy scales is responsible for 
the double-peak structure of the temperature dependence of the 
specific heat.  

\begin{figure}[thb]
\begin{center}
\resizebox{7.0cm}{!}{\includegraphics{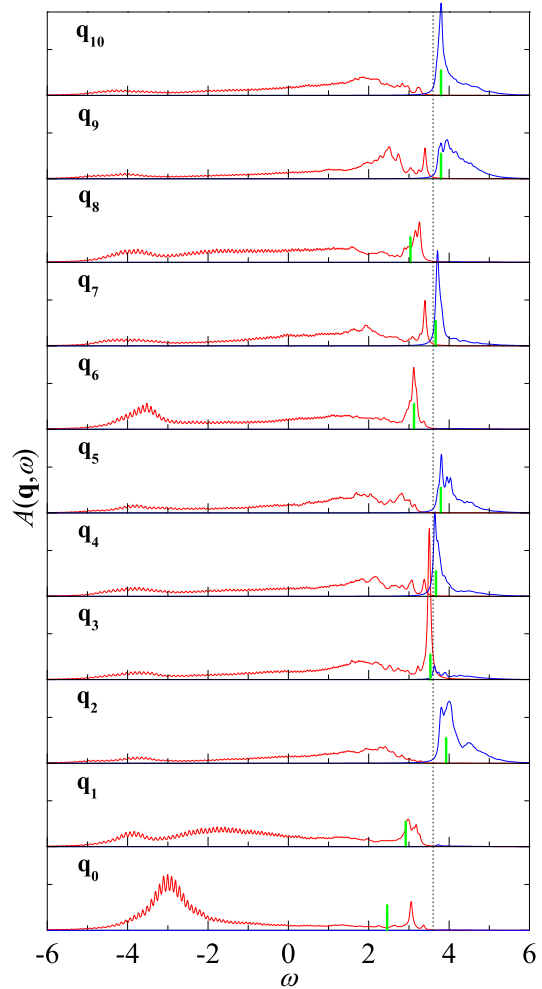}}\\
\caption{(Color online) Calculated single-particle excitation 
spectra of the $t$-$J$-$K$ model for the ground state of the 
20-site cluster with two vacancies ($n=0.9$).  Vertical dotted 
line represents the Fermi energy.  Vertical bars represent the 
position of the noninteracting band dispersion with a reduced 
hopping parameter $t_{\rm eff}=t/6$.  
We assume $J=-0.3$, $K=0.06$, and $\eta=0.04$}
\label{singleparticlespectra}
\end{center}
\end{figure}

\subsection{Single-particle excitation spectra}

The calculated results for the single-particle excitation 
spectra $A({\bf q},\omega)$ are shown in Fig.~6.  From the 
results, we can deduce the possible quasiparticle band 
structure and hence the Fermi-surface topology.  
We find the following.  
There are broad and incoherent spectral features over a wide 
energy range corresponding to the total band width of the 
noninteracting dispersion, 
i.e., $-5\lesssim\omega\lesssim 5$, but there emerge the 
sharp quasiparticle-like peaks with a characteristic 
dispersion in the vicinity of the Fermi energy.  This result 
is similar to the case of the square-lattice $t$-$J$ model 
near half filling.\cite{Eder1,Eder2}  
Let us assume this to be the consequence of the presence 
of fermionic quasiparticles.  Then, we find the quasiparticle 
band structure to be fitted well by the noninteracting band 
dispersion with a reduced hopping parameter $t_{\rm eff}$ 
or with an enhanced effective mass $m^*$ of the quasiparticle 
(see Fig.~7).  From the fitting, we find the value 
\begin{equation}
t_{\rm eff}\,\,=\,\,\frac{m}{m^*}\,t\,\,\simeq\,\,(1/6)t
\end{equation}
or $m^*/m\simeq 6$ at $n=0.9$ and $K=0.06$.  
The present results also suggests that the Fermi-surface 
topology of the quasiparticles is equivalent to that of 
the noninteracting system since the quasiparticle band can 
be obtained only by assuming that the band width is reduced 
(or the band mass is enhanced).  
Thus, the Fermi surface is large (i.e., its area $\propto n$) 
rather than small (i.e., its area $\propto(1-n)$).  
The doping dependence of $m^*$ should be interesting, 
in particular whether $m^*$ diverges or not 
at $n\rightarrow 1$.  However, we cannot answer this 
question in our small-cluster study; the behavior of $m^*$ 
even in the square-lattice $t$-$J$ model still remains to 
be a puzzle.  

\begin{figure}[t]
\begin{center}
\resizebox{8.5cm}{!}{\includegraphics{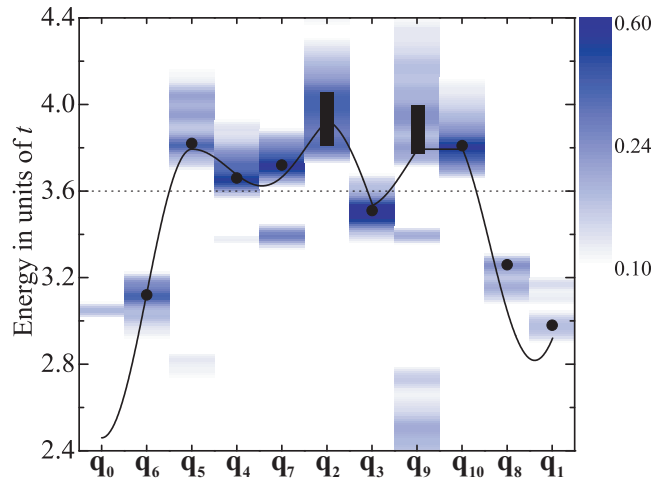}}\\
\caption{(Color online) Quasiparticle band structure of the 
$t$-$J$-$K$ model.  Darkness of the shadow is in proportion 
to the spectral weight $A({\bf q},\omega)$ shown in Fig.~6.  
Solid circles and bars represent the peak positions of the 
single-particle spectra.  The peak positions are fitted with 
the noninteracting band structure (solid line) with a reduced 
hopping parameter $t_{\rm eff}=t/6$.  Horizontal line is 
the Fermi energy.}
\label{quasiparticledispersion}
\end{center}
\end{figure}

We may also assume that the quasiparticle band width shown in 
Fig.~7 may scale well with the energy of the spin excitations, 
i.e., a combination of the exchange parameters $J$ and $K$, 
while the entire band width ($\sim 9t$) of the broad spectral 
features shown in Fig.~6 scales with $t$, as in the case of the 
square-lattice $t$-$J$ model near half filling.\cite{Eder1,Eder2}  
Thus, we again find the separation between the spin and density 
degrees of freedom in their energy scales.  
The quasiholes (or quasiparticles as their conjugate) are 
thus the vacancies dressed by the spin excitations.  
We should note therefore that the spin and density degrees 
of freedom are not exactly separated in this sense, unlike in 
the Tomonaga-Luttinger liquid in the one-dimensional interacting 
fermion systems.\cite{TLL}  Only the energy scales are different.  
Further experimental and theoretical studies will be required 
to clarify the true low-energy physics of the system.  

We may also point out that the enhanced effective mass of the 
quasiparticle band structure may partly be responsible for 
the enhancement of the effective mass $m^*/m$ determined from 
the specific heat coefficient $\gamma$ although the latter 
cannot be obtained from our finite-size calculations.  
In Sec.~IV, we compare the effective mass obtained from the 
quasiparticle band dispersion with experiment.  

\begin{figure}[th]
\begin{center}
\resizebox{8.5cm}{!}{\includegraphics{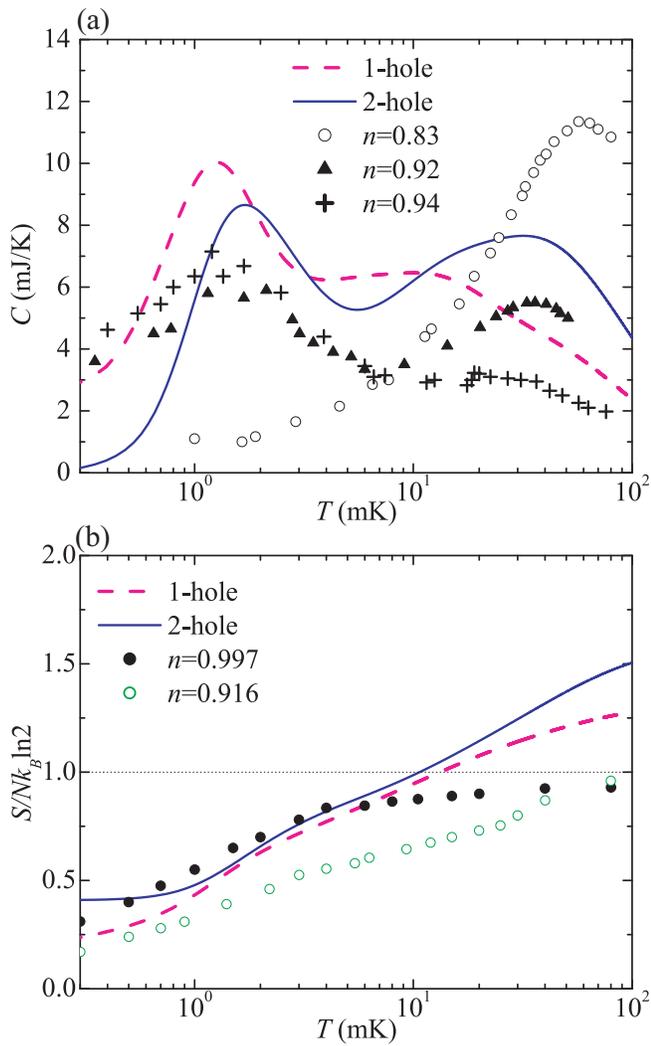}}\\
\caption{(Color online) Temperature dependence of 
(a) the heat capacity $C(T)$ in units of mJ/K and 
(b) entropy $S(T)$ divided by the entropy of $N$ free 
spins $Nk_{\rm B}\ln 2$.  
Comparisons between theory (solid and dashed lines) 
and experiment (symbols) are shown.  Experimental data 
are taken from Refs.~\cite{Matsumoto1,Matsumoto2}.}
\label{exptCandS}
\end{center}
\end{figure}

\section{Comparison with experiment}

The calculated results for the heat capacity $C(T)$ at 
$J=-0.3$ and $K=0.06$ (see Sec.~III B) at the fillings 
of $n=0.92$ and $0.83$ are compared with experiment 
in Fig.~8(a).  We here assume the value of $t$ determined 
so as to reproduce the higher-temperature peak observed 
in $C(T)$ at $n=0.92$, i.e., $t=43.8$ mK, so that we have 
$|J|=13.1$ mK and $K=2.63$ mK with keeping the ratio 
$t$ : $|J|$ : $K$ = $1 : 0.3 : 0.06$.  
We also assume that the total area of the sample used in 
experiment (556 m$^2$) is uniformly active and contributes 
to the heat capacity.  
We should note that the results for the temperature 
region $T\lesssim 1$ mK are not reliable because of 
the finite-size effects where the discreteness of the 
energies in the system gives the exponential decay of 
the heat capacity at low temperatures.  

We then find the fair agreement with experiment; 
in particular, the double-peak structure in $C(T)$, i.e., 
the lower-temperature peak that comes from the spin 
excitations and higher-energy peak that comes from the 
density excitations of the system, are reasonably 
well reproduced.  
More precisely, we find that our calculated results 
reproduce the experimental tendency that, near half filling, 
the lower-temperature peak is high but with increasing 
the vacancy concentration, the higher-temperature peak 
becomes larger and simultaneously the peaks shift to 
higher temperatures.  

Note that the specific heat coefficient $\gamma$ (or the 
effective mass) cannot be estimated from the present 
calculations of $C(T)$ due to finite-size effects.  
However, we find that the value of the enhanced effective 
mass $m^*/m\simeq 6$ estimated from the calculated quasiparticle 
band structure (see Sec.~III E) is consistent with the 
experimental value $\sim 7.5-10$ estimated from the observed 
temperature dependence of $C(T)$ at $n=0.89$.\cite{Matsumoto2}
Here, we should note that the definition of $m^*$ in the 
experimental specific heat coefficient is two-fold: 
one is the value deduced from the lower-temperature peak 
and the other is the value deduced from the higher-temperature 
peak.  The two values are, however, not very different at 
least for $n\lesssim 0.9$, so that we can make comparison 
with our theoretical value.  
Then, it seems reasonable to assume that the 
renormalization of the band structure due to the spin 
excitations is mainly responsible for the observed 
enhancement of the effective mass.\cite{Matsumoto2}  
Thus, stated differently, the specific heat coefficient 
$\gamma$ should be determined predominantly by the spin 
excitations of the system.  

The calculated results for the entropy $S(T)$ are also 
compared with experimentally determined\cite{Matsumoto2} 
entropy in Fig.~8(b).  We again find the fair agreement in 
their general tendencies.  
More precisely, we find that the calculated curves of $C(T)$ 
cross the line of the entropy of $N$ free spins 
$Nk_{\rm B}\ln 2$ at $\sim$10 mK, at which the 
lower-temperature peak in $C(T)$ terminates.  This result also 
supports that the lower-temperature peak in $C(T)$ comes from 
the excitations of the spin degrees of freedom of the system.  
The higher-energy peak should therefore come from the 
density degrees of freedom or motions of vacancies in the 
system.  
Note that the experimentally determined\cite{Matsumoto2} 
entropy is significantly smaller than $Nk_{\rm B}\ln 2$ even 
in the vicinity of half filling, $n=0.997$, and even at 
temperatures of $10-20$ mK where the lower-temperature 
peak in $C(T)$ terminates.  
The missing entropy may reside in the region of much 
lower temperatures that the present experiment does not 
approach.\cite{Fukuyama2}  

\begin{figure}[th]
\begin{center}
\resizebox{7.0cm}{!}{\includegraphics{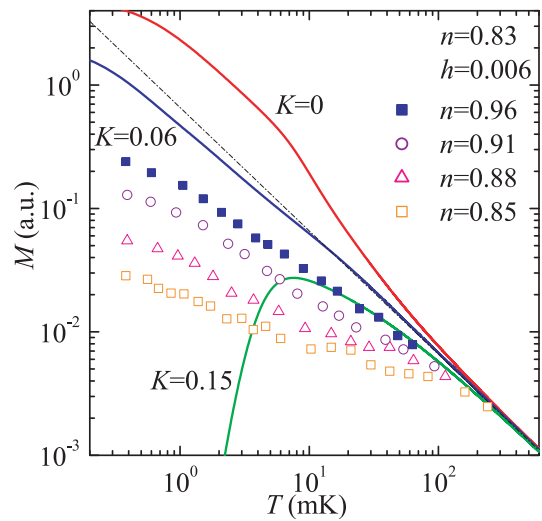}}\\
\caption{(Color online) Temperature dependence of the 
magnetization $M$ under the uniform magnetic field $h$.  
Comparison between theory (solid lines) and experiment 
(symbols) is shown.  The straight line corresponds to 
the Curie law.  Experimental data are taken from 
Ref.~\cite{Murakawa}.}
\label{exptCHI}
\end{center}
\end{figure}

The calculated results for the magnetization under the 
uniform magnetic field are compared with experiment in 
Fig.~9.  We should note that the experimentally applied 
magnetic field, which is $h=0.006$ in our calculations, 
is very small in comparison with the energy scales of 
the $^3$He system, so that the behavior of the 
magnetization under the uniform magnetic field is the 
same as that of the uniform magnetic susceptibility 
defined at $h\rightarrow 0$.  

We find that, although the finite-size effect is strong 
at low temperatures, the calculated magnetization at 
$K=0.06$ is consistent with experiment in the sense 
that the value is somewhat smaller than the value 
expected from the Curie law $M(T)=\chi(T)h=Ch/T$.  For more 
quantitative comparison, however, one would need the techniques 
appropriate for treating infinite-size systems, so that 
experimentally observed plateau-like behavior\cite{Murakawa} 
in the temperature dependence of the magnetization 
can be explained.  

\section{Summary}

We have used an exact-diagonalization technique on small 
clusters to study the low-energy physics of the 
triangular-lattice $t$-$J$ model with the multiple-spin 
exchange interactions, whereby we have considered the 
anomalous properties observed in the doped Mott region of 
the two-dimensional liquid $^3$He adsorbed on a graphite 
surface.  
We have calculated the temperature dependence of the 
specific heat, entropy, and uniform magnetic susceptibility, 
as well as the spin and density excitation spectra and 
single-particle spectra for the model, and have considered 
their implications.  

We have shown the following: 

\noindent
(1) The double-peak structure appears in the temperature 
dependence of the specific heat.  The result is quantitatively 
consistent with experiment.  The low-temperature sharp peak 
comes from the spin excitations and high-temperature broad 
peak comes from the density excitations.  

\noindent
(2) The spectral weight for the spin excitations is 
concentrated on a very low-energy region, the width of which 
scales with the exchange interactions, while that of the 
density excitations extends over an entire band width, which 
scales with the hopping parameter of the vacancy.  The clear 
separation between spin and density excitations in their energy 
scales is thus found.  

\noindent
(3) The accumulation of the spectral weight of the spin 
excitations comes from the frustrated nature of the spin 
degrees of freedom of the system; i.e., the ferromagnetic 
two-spin interactions $J$ compete with the antiferromagnetic 
four-spin interactions $K$ on the geometrically frustrated 
triangular lattice.  

\noindent
(4) The single-particle excitation spectra suggest that the 
vacancies behave like the fermionic quasiparticles dressed 
by the spin excitations, of which the effective band mass is 
estimated to be $m^*/m\simeq 6$ at $n=0.9$, in consistent 
with the effective mass measured from the specific heat 
coefficient.  

\noindent
(5) The temperature dependence of the spin susceptibility 
shows a suppressed Curie-like behavior, reflecting the 
situation where the ground state is highly degenerate due 
to the frustrated nature of the spin degrees of freedom.  

We hope that the present study will shed more light on 
the physics of the two-dimensional $^3$He systems and 
stimulate further experimental and theoretical studies of 
the systems in greater details.  
We have focused on the doped Mott region of the monolayer 
$^3$He in this paper.  However, it has recently been 
reported\cite{Neumann} that the bilayer $^3$He systems 
also contain rich physics concerning heavy fermions with 
quantum criticality, which we want to leave for future 
study.  

\begin{acknowledgments}
Enlightening discussions with Professors Hiroshi Fukuyama 
and John Saunders are gratefully acknowledged.  
This work was supported in part by Grants-in-Aid for 
Scientific Research (Nos.~18028008, 18043006, 18540338, 
and 19014004) from the Ministry of Education, Culture, 
Sports, Science and Technology of Japan.  
TS acknowledges financial support from JSPS Research 
Fellowship for Young Scientists.  
A part of computations was carried out at the 
Research Center for Computational Science, 
Okazaki Research Facilities, and the Institute 
for Solid State Physics, University of Tokyo.  
\end{acknowledgments}

\end{document}